# The hidden force opposing ice compression[1]

Chang Q Sun,[1,2,3*] Xi Zhang,[1] Weitao Zheng[2]

[1] *School of Electrical and Electronic Engineering, Nanyang Technological University, Singapore 639798*

[2] *Department of materials Science, Jilin University, Changchun Changchun 130012, China*

[3] *Faculty of Materials and Optoelectronic Physics, Xiangtan University, Hunan 411105, China*

*E-mail: Ecqsun@ntu.edu.sg;*

Abstract

Coulomb repulsion between the unevenly-bound bonding "-" and nonbonding ":" electron pairs in the "$O^{2-}$ : $H^{+/p}$-$O^{2-}$" hydrogen-bond is shown to originate the anomalies of ice under compression. Consistency between experimental observations, density functional theory and molecular dynamics calculations confirmed that the resultant force of the compression, the repulsion, and the recovery of electron-pair dislocations differentiates ice from other materials in response to pressure. The compression shortens and strengthens the longer-and-softer intermolecular "$O^{2-}$ : $H^{+/p}$" lone-pair virtual-bond; the repulsion pushes the bonding electron pair away from the $H^{+/p}$ and hence lengthens and weakens the intramolecular "$H^{+/p}$-$O^{2-}$" real-bond. The virtual-bond compression and the real-bond elongation symmetrize the "$O^{2-}$-$H^{+/p}$ : $O^{2-}$" as observed at ~60 GPa and result in the abnormally low compressibility of ice. The virtual-bond stretching phonons (< 400 cm$^{-1}$) are thus stiffened and the real-bond stretching phonons (> 3000 cm$^{-1}$) softened upon compression. The cohesive energy of the real-bond dominates and its loss lowers the critical temperature for the VIII-VII phase transition. The polarization of the lone electron pairs and the entrapment of the bonding electron pairs by compression expand the band gap consequently. Findings should form striking impact to understanding the physical anomalies of $H_2O$.

---

[1] This presentation is associated with supporting information.

# Contents

# I     Introduction

$H_2O$ has been the subject of extensive study, given its paramount importance in nature science [1, 2, 3, 4, 5] and its role in DNA folding [6, 7], protein and gene delivery [8, 9]. Considerable achievements have been made in past decades towards: i) optimization of crystal structures, phase formation and transition, and the binding energy under various conditions [10, 11, 12, 13]; ii) understanding the reaction dynamics of $H_2O$ with other ingredients [14, 15]; iii) quantification of the hydrogen-bond weak force interactions [16, 17]; . The up-to-date knowledge of ice under compression includes: i) ice turns to be partially ionic at extremely high pressure (2 TPa) and high temperature (2000 K) [13]; ii) the contribution to the lattice energy from the van der Walls intermolecular interaction (nonbonding lone pair named herewith) increases and that from the intramolecular hydrogen bonding (the real bond) decreases when pressure is increased up to 2 GPa, as calculated using first principle and Quantum Mote Caro calculations [18]. The elegantly used TIPnP (n varies from 1 to 5) model series and the TIP4Q/2005 modeled water ice but they exclude the possibility of bond angle and length relaxation and charge polarization. These rigid non-polarizable models can hardly reproduce the anomalies of water ice with high satisfaction [19, 20].

As indicated by Ball [1], water ice is too strange, too anomalous, and too challenge. Clarification of their physical origins and theoretical reproduction of the measured anomalies remains a great challenge [21, 22, 23, 24, 25, 26] For instances, it is usual for other materials that the critical temperature ($T_C$) for liquid-solid or disordered-ordered phase transition increases with the applied pressure ($P$) in a quasi-equilibrium process [27, 28]; however, the $T_C$ for ice transferring from ice-VIII phase to the proton-disordered ice-VII drops from 280 to 150 K when the P is increased from 1 to 50 GPa [29]. Compression shortens the O---O distance but lengthens the O-H bond, leading to the low-compressibility and the proton symmetrization of ice-VIII at about 59 GPa and 0.20 nm O---O distance [5, 30]. Generally, the applied pressure stiffens all the Raman phonons of other materials such as carbon allotropes [31]; however, the ice-VIII vibration spectra are opposite; the softer Raman modes at frequency lower than 400 $cm^{-1}$ are stiffened but the stiffer mode at frequency greater than 3000 $cm^{-1}$ are softened [4, 29, 32]. These discoveries inspired us seeking for the hidden force driving the discovered anomalies and reproducing quantitatively the observations.

The aim of this communication is to show that a segmentation of the "$O^{2-}$-$H^{+/p}$ : $O^{2-}$" hydrogen bond,

with the aid of density functional theory (DFT) and molecular dynamics (MD) calculations, has enabled us to clarify and correlate these concerns with improved understanding of their common origin. Being able to quantitatively reproduce the anomalies of proton symmetrization, phonon relaxation, volumetric and $T_C$ anomalies of frozen $H_2O$ under compression, we uncovered that Coulomb repulsion between the unevenly-bound bonding "-" and nonbonding ":" electron pairs in the "$O^{2-}$ : $H^{+/p}$-$O^{2-}$" hydrogen bond forms the key to the anomalies of concern.

**II       Hypothesis and expectations: Repulsion between the electron pairs**

Instead of the widely-used rigid non-polarizable models (as compared in the supporting information) [19, 20], we show in Figure 1 the segmentation of the "$O^{2-}$ : $H^{+/p}$-$O^{2-}$" hydrogen bond, as the basic structural unit in ice and water, into the longer-and-weaker "$O^{2-}$ : $H^{+/p}$" intermolecular virtual-bond and the shorter-and-stronger "$H^{+/p}$-$O^{2-}$" intramolecular real bond. As the coordinate origin, the $H^{+/p}$ plays a dual role of $H^+$ and $H^P$. The $H^{+/p}$ donates its electron to one $O^{2-}$ to form the real-bond and meanwhile it is polarized by the nonbonding lone pair of the other neighboring $O^{2-}$ upon the sp-orbit of oxygen being hybridized in reaction [33]. This segmentation shows the non-rigid and polarizable nature of the H-bond, instead, which is the key to the H-bond asymmetric relaxation dynamics under external stimuli.

In the hexagonal or cubic ice, the $O^{2-}\cdots O^{2-}$ distance is 0.276 nm. The intramolecular $H^{+/p}$-$O^{2-}$ real bond is much shorter and stronger (~0.100 nm and ~$10^0$ eV) than that of the intermolecular $O^{2-}$ : $H^{+/p}$ nonbond (~0.176 nm and ~$10^{-2}$ eV). The angle between the $H^{+/p}$–$O^{2-}$–$H^{+/p}$ is smaller than 104.5° while the angle between the $H^{+/p}$ : $O^{2-}$ : $H^{+/p}$ is greater than 109.5° for a free $H_2O$ molecule. The $O^{2-}$ : $H^{+/p}$-$O^{2-}$ deviates from the $O^{2-}\cdots O^{2-}$ line by only several degrees in ideal case that is negligible.

The $O^{2-}$ : $H^{+/p}$-$O^{2-}$ represents the average of all the hydrogen bonds involved in water ice unless at extreme conditions [13]. This average minimizes the fluctuations in the number, topological, length scale, boundary, etc [34, 35]. The surrounding interactions by other $H_2O$ molecules units are also averaged as the background. At extremely low proton mass and low temperature, quantum effect may come into play, but this effect may cause fluctuation. In fact, the fluctuation only affects the precision of the derived information but the nature of the observations. The advantage of such an extended Ice Rule is that it allows us to focus on the responses of the segments to applied stimulus separately and their corporative interaction.

Our hypothesis is that the $O^{2-}:H^{+/p}$ and the $O^{2-}-H^{+/p}$ response to the pressure not in the same way but one serves as the master and the other a slave. The master $O^{2-}:H^{+/p}$ virtual-bond is readily compressed because it is much softer than the real bond that serves as a slave. As denoted in Figure 1, the resultant force of the compression $f_p$, the Coulomb repulsion $f_q$, and the recovery of electron-pair dislocation $f_r$ determines the extents of $O^{2-}$ displacements and hence the lengths and strengths of the real and the virtual bond segment- asymmetric relaxation.

It is expected that under compression, the $O^{2-}:H^{+/p}$ becomes shorter and stronger, the ":" will move towards the $H^{+/p}$ and push the bonding pair slightly away from the $H^{+/p}$ origin. The $H^{+/p}-O^{2-}$ bond then becomes longer-and-weaker but the result $O^{2-}\cdots O^{2-}$ becomes shorter. It is also expected that such a process of asymmetric relaxation symmetrizes the "$O^{2-}-H^{+/p}:O^{2-}$" and results in the unusually low compressibility of ice. The virtual-bond stretching phonons ($< 400$ cm$^{-1}$) will shift to higher frequencies and the real-bond stretching phonons ($> 3000$ cm$^{-1}$) will be softened upon compression because the frequency shift is proportional to the stiffness of the segmented interactions. The cohesive energy loss of the real-bond dominates and lowers the critical temperature for the VIII-VII phase transition as the binding energy of the virtual-bond is negligibly small; the polarization of the lone electron pairs and the entrapment of bonding electrons upon compression will expand the band gap consequently.

In order to verify the hypotheses and expectations, we conducted the MD and DFT calculations and Raman spectroscopic analysis. Details of the calculations are described in the methodology section. The numerical convergence and the DFT dispersions are given in the supporting information.

**III     Results and discussion**

*3.1     Proton symmetrization and the low compressibility*

Figure 2(a) shows the pressure-induced relaxation dynamics of the virtual and real bond segments. The MD results show that the $H^{+/p}:O^{2-}$ nonbond is compressed from 0.1767 to 0.1692 nm and meanwhile the $H^{+/p}-O^{2-}$ bond is elongated from 0.0974 to 0.1003 nm when the pressure is increased from 1 to 20

GPa. The relaxation of each segment, denoted with subscript x, can be represented using the polynomial form, $d_x = d_{x0}[1 + \alpha_x(P - P_0) + \beta_x(P - P_0)^2]$ with $P_0 = 1$ GPa. Encouragingly, the calculated O-H and O : H distances agree exceedingly well with the trends reported in refs [5, 36, 37]. The DFT outcome also shows the same trend despite the deviation in slopes at low pressures. The DFT deviation may arise from the artifacts involved in the ab initio algorithm optimization. As it will be shown, the MD results meet the constraints (eq 1) for the H-bond asymmetric relaxation dynamics.

It is exciting that, as shown in Figure 2(a), an extrapolation of the MD-derived polynomial expressions leads to the proton symmetrization occurring at 58.6 GPa with the $O^{2-}$---$O^{2-}$ distance of 0.221 nm, which is in good accordance with the reported values of 59 GPa and 0.220 nm[30]. In 1972, Holzapfel [38] predicted that, under pressure, hydrogen bonds might be transformed from the highly asymmetric $O^{2-}$ - $H^{+/p}$ : $O^{2-}$ configuration to a symmetric state in which the $H^+$ proton lies midway between the two $O^{2-}$ ions, leading to a non-molecular symmetric phase of ice, which contradicts clearly with the rigid non-polarizable models. This prediction was numerically confirmed in 1998 by Benoit, Marx, and Parrinello [30] who proposed that the "translational proton quantum tunneling under compression" dominates this phenomenon. In the same year, an *in situ* high-pressure Raman measurement conducted by Goncharov et al [39] confirmed that the proton symmetry happens at 60 GPa and 100 K, as no further phonon relaxation could be observed with the increase of pressure.

Yoshimura et al [4] firstly reported a comprehensive set of V-P data of ice-VIII measured using the *in situ* high-pressure and low-temperature synchrotron x-ray diffraction and Raman spectroscopy. This set of data was reproduced using the current MD and DFT calculations, as shown in Figure 2(b). The matching to the MD outcome gives the state of equation, $V/V_0 = 1 - 2.38 \times 10^{-2} P + 4.70 \times 10^{-4} P^2$ with $V_0 = 1.06$ cm$^3$/kg.

Consistency between the MD and DFT-derived and the reported proton symmetrization [30, 39] and low compressibility [4] of ice verified our hypothesis that the weaker lone pair is highly compressed yet the stronger bonding pair is elongated because of the resultant force of the compression, the repulsion, and the recovery for dislocations of the unevenly-bound electron pairs. The hidden repulsion between the unevenly-bounded electron pairs forms indeed the key to the observed symmetric and volumetric anomalies of ice.

*3.2    The slopes and curvatures of the relaxation curves*

According to the configuration in Figure 1 and the findings in Figure 2, we can correlate the force constants, $k_H$ and $k_L$, of the segments in the hydrogen bond. For each electron pair, there are three forces being acted on, i.e., the compression force $f_p \sim P/s_x$, the repulsion force $f_q \sim (d_{O\text{-}O})^{-2}$, and the deformation recovery force $f_r = -k_H \delta d_H$ or $-k_L \delta d_L$ opposing to the dislocation direction. The P and s represent the identical pressure and the disparic cross section of the real and the virtual bond. The resultant of the three forces determines the equilibrium of the electron pairs in the hydrogen bond. The $\delta d_H$ and $\delta d_L$ are the dislocations of the respective electron pairs.

One can readily derive that the equilibrium of these forces leads to the criterion of volume compression,

$$f_{pL} - f_{pH} = P(1/s_L - 1/s_H) = (f_{rL} + f_{rH}) > 0, \tag{1}$$

This relationship, as the constraint for the H-bond asymmetric relaxation dynamics, indicates that the effective cross section area of the virtual bond is smaller than that of the real bond. From the relaxation trends in Figure 2(a), one can see that the slope and the curvature of the $d_x$-P curve are opposite, which means that if one segment contracts the other lengthens. The slopes and the curvatures of the dislocation curves of the segments are always opposite in sign, as shown in Figure 2(a). From this constraint perspective and the calculated trends in Figure 2(a), the MD calculation is preferred in this regard.

*3.3    Raman phonon relaxation*

Figure 3(a) shows the MD-derived power spectra in the specified frequency ranges of $\omega_L <$ 400 cm$^{-1}$ and $\omega_H >$ 3000 cm$^{-1}$ as a function of pressure. As P increases, the $\omega_H$ is softened from 3520 cm$^{-1}$ to 3320 cm$^{-1}$ and the $\omega_L$ is stiffened from 120 to 336 cm$^{-1}$, disregarding the possible phase change and other supplementary peaks nearby. As compared in Figure 3(b), the currently MD-derived phonon relaxation trends are in good agreement with the trends of ice-VIII measured at 80 K using Raman [4, 29] and infrared (IR) spectroscopies [32, 40]. The consistency between calculations and measurements of the branched phonon relaxation dynamics confirms our expectations. As we focus on the nature, the origin, and the trend of change, the slight deviation caused by artifacts or errors in measurements and calculations is within the tolerance.

The Raman spectroscopy is one of the powerful techniques that could discriminate the vibrations of the inter-molecular virtual bond and the intromolecular real bond in different frequencies, according to their stiffness. From the first-order approximation, the vibrations of the two segments of the H-bond can be taken as a harmonic system each with an interaction potential, $u_x(r)$. Equaling the vibration energy of the harmonic system to the third term of the Taylor series of its interaction potential at equilibrium, we can obtain the relation:[41, 42]

$$\frac{1}{2}\mu(\Delta\omega)^2(r-d_x)^2 = \frac{1}{2}k_x(r-d_x)^2 \cong \frac{1}{2}\frac{\partial u(r)}{\partial r^2}\bigg|_{r=d_x} x^2$$

$$\propto \frac{1}{2}\frac{E_x}{\mu d_x^2}(r-d_x)^2$$

$$\Delta\omega_x = \omega_x - \omega_{x0} \propto \frac{E_x^{1/2}}{d_x}$$

(2)

The $k_x$ is the force constant of the x branch at equilibrium. The Raman shift $\Delta\omega_x$ depends on the length and energy of the bond and the reduced mass µ of the atoms or molecules of the dimer vibronic system. From the dimensional point of view, the second order derivative of the $u_x(r)$ at equilibrium is proportional to the bind energy $E_x$ divided by the square bond length in the form of $d_x^2$. The $E_x^{1/2}/d_x \cong \sqrt{Y_x d_x}; (Y_x \approx E_x/d_x^3)$ is right the square root of the stiffness being the product of the Young's modulus and the bond length [40]. Therefore, the Raman shift is proportional to the square-root of the bond stiffness.

The intrinsic vibration frequency of the bond is detectable as the Raman shift from the referential point $\omega_{x0}$. This relationship indicates that a blue shift will happen if the bond is stiffened, or the bond length is shortened or the binding energy is increased, and vice versa. Generally, the bond energy is inversely proportional to the bond length in a certain power [40]. Therefore, the frequency shift of the soft and the stiff Raman mode fingerprints the change of the length and energy of the respective segment of the hydrogen bond. The facts of the soft-phonon stiffening and the stiff-phonon softening confirm that the longer-and-weaker lone-pair virtual-bond becomes shorter-and-stiffer while the shorter-and-stronger real-bond becomes longer-and-softer, as shown in Figure 2. This derivative is also consistent with the latest discovery[18] that the intermolecular lone pair interaction increases and the intromolecular bonding energy decreases as the pressure is increased.

## 3.4 P-depressed critical temperature of phase transition

The correlation between the bond length ($d_x$), bond energy ($E_x$), and the $T_C$ of a system under compression has been established as follows according to the local bond average approach,[27, 40]

$$\frac{T_C(P)}{T_C(P_0)} = \frac{E_x(P)}{E_x(P_0)} = 1 + \Delta_p = 1 - \frac{\int_{V_0}^{V} p(v) dv}{E_x(P_0)} = 1 - \frac{\int_{P_0}^{P} p \frac{d d_x}{d p} d p}{E_x(P_0)/V_{x0}}$$

$$\frac{d d_x}{d p} = d_{x0}[\alpha_x + 2\beta_x(P - P_0)]$$

(3)

Generally, the $T_C$ is proportional to the atomic cohesive energy [40]. Because of the "isolation" of the $H_2O$ molecule by the nonbonding lone pairs, the $T_C$ of ice should be proportional to the cohesive energy of the real bond in a $H_2O$ molecule. $E_x(P_0)/V_{x0}$ is the binding energy density per bond with the given volume $V_{x0}$.

Figure 4 shows the consistency between calculated using eq (3) and the measured $T_C$ for the VII-VIII phase transition. The consistency in the $T_C$-$P$ trend justifies that the change of $T_C$ is indeed dominated by the binding energy of the real bond, as the binding energy of the lone pair in the $10^{-2}$ eV level is negligibly small. From matching to the measured $T_C$-$P$ curve of ice transferring from VIII to VII phase [29, 32, 43], we estimated the binding energy of the real-bond as $E_{HO}$ (at 1 GPa) = 3.97 eV by assuming the real bond diameter as that of the atom [44]. For H, it is 0.0.53 nm.

## 3.5 Band gap expansion

Figure 5 shows the DFT-derived energy dispersion and the evolution of the density of states (DOS) of ice-VIII with pressure varying from 1 to 60 GPa (see full datum in supporting information). The DOS above the Fermi level originates from the polarization states of the lone pair and the valence DOS is dominated by the bonding states of oxygen [33, 45].

It is seen from Figure 5(a) that the bottom edge of the valence band shifts down from −6.7 eV at 1 GPa to −9.2 eV at 60 GPa; while the conduction band shifts up from 5.0~12.7eV at 1 GPa to 7.4~15.0eV at 60 GPa. This band gap enlargement arises from the polarization of the lone pair and the entrapment of the bonding electrons by compression. The bandgap expands further at higher pressure from 4.5 to 6.6 eV, as shown in Figure 5(b) when the P is increased from 1 to 60 GPa.

## IV Conclusion

Quantitative matching between the MD-DFT calculations and experimental observations verified our hypothesis and the associated expectations of ice under compression. We may conclude:

1. Coulomb repulsion between the unevenly-bonded electron pairs and the discriminative response of the real and the virtual bond to the applied stimuli form the key to the physical anomalies of frozen $H_2O$ upon compression.
2. The resultant forces associated with compression, Coulomb repulsion, and the recovery of dislocation of the electron pairs in the respective segment of the hydrogen bond dominates the *P*-derived proton symmetrization, vibration, volume compression, and phase transitional anomalies of ice under compression.
3. The initially longer-and-weaker nonbond becomes shorter-and-stiffer but the initially shorter-and-stronger real-bond becomes longer-and-softer under increased pressure, which is in line with the latest findings of [18] energy contribution of the segments to the lattice energy.
4. The initially softer phonons are stiffened and the stiffer phonons are softened as consequence of the asymmetric relaxation of the real and virtual bond segment.
5. The pressure-enhanced band gap expansion evidences the polarization of the lone pair electrons and entrapment of the bonding electrons upon compression.
6. A segmentation of the hydrogen bond and the asymmetric relaxation and polarizablility are necessary to examine the length, energy, vibration frequency response of each to external stimuli and their consequence on macroscopic properties of ice.

**Methodology: MD and DFT numerical computations**

The MD calculations were performed using Forcite's package with *ab initio* optimized forcefield Compass27 [46]. The Compass27 has been widely used in dealing with the electronic structures and the hydrogen bond network of water and amorphous ices [47] as well as water chains in hydrophobic crystal channels [48].

DFT calculations were conducted on ice-VIII unit cell by using the CASTEP code[49] within the Perdew-Burke-Ernzerhof functional (PBE)[50] functional parameterization of generalized gradient approximation. Norm-conserving pseudopotential (NCP) was adopted where the $1s^1$ and the $2s^22p^4$ are treated as valence electrons for H and O atoms, respectively. The use of the plane-wave kinetic energy cutoff of 500 eV adopted here, were shown to give excellent convergence of total energies. During calculations, the self-consistency threshold of total energy was set at $10^{-6}$ eV/atom. In geometry optimizations of ice-VIII from 1 to 60 GPa, the tolerance limit for the energy, forces and displacement were set at $10^{-5}$ eV/atom, 0.03 eV and 0.001 Å, respectively.

Ice-VIII consisting of two interpenetrating cubic ice lattices, of 8 molecules in each unit cell is examined. The MD calculations were performed to examine the evolution of the O-H and O : H distances in a 2×2 supercell of ice-VIII unit, containing 32 molecules, under the pressure changing from 1 to 20 GPa. The structure was dynamically relaxed under the Isoenthalpic–isobaric ensemble for 30 ps, showing sufficiently stable convergence (supporting information). The average O-H and O : H lengths were taken of the structures of the last 10 ps (20,000 steps). The power spectra were calculated from the Fourier transformation of their velocity autocorrelation function, Cor(t) [51], $I(\omega) = 2\int_0^\infty \text{Cor}(v(t))\cos\omega t \, dt$ using velocity data of all atoms recorded at every 0.5 fs.

Table and figure captions

Figure 1 The segmentation of the commonly-known "$O^{2-}$ : $H^{+/p}$-$O^{2-}$ hydrogen-bond with notion of the Coulomb repulsion force $f_q$, the compression force $f_p$, and the dislocation recovery force $f_{rx} = -k_x \Delta d_x$. This framework correlates the force constants of the lone pair ($k_L$) and the bonding pair ($k_H$) as inversely proportional to the slope and the curvature of the respective $d_x$-P curve:

$$\frac{k_H}{k_L} = -\frac{\delta d_L/\delta p}{\delta d_H/\delta p} = -\frac{\delta^2 d_L/\delta p^2}{\delta^2 d_H/\delta p^2} .$$

Figure 2 (a) MD and DFT derivatives of the pressure-induced ":" and "–" asymmetric relaxation dynamics and the proton centralization occurring under 58.6 GPa compression at a O-O distance of 0.221 nm, are in exceedingly good accordance with the reported results (EXP) at 59 GPa and 0.22 nm [30, 39]. (b) Matching of the MD and DFT-derivatives to the measured [4] V-P curve of ice.

Figure 3 (a) MD-derived power spectra of ice-VIII under pressure in comparison to (b) the measured vibration peaks of $\omega_H$(circle) and $\omega_\Lambda$(square) of ice-VIII at 80 K[4], using Raman[29] and infrared absorption [32]. The trends consistency evidences that the $H^{+/p}$ : $O^{2-}$ is shortened and strengthened (blue shift of the low-frequency mode <400 cm$^{-1}$) and the $H^{+/p}$-$O^{2-}$ is eelongated and weakened (red shift of the high-frequency stretching mode >3000 cm$^{-1}$) [40].

Figure 4 Consistency between the theoretically (eq 3) derived and the measured (Exp-1, Exp-2) $T_c$-P for ice VIII-VII transition [29, 32, 43]. The real-bond energy dictate the $T_C$ as the contribution from the virtual-bond energy is negligibly small.

Figure 5 (a) DFT-derived DOS of ice-VIII at 1, 20, 40 and 60 GPa. As pressure increases, the bottom edge of the valence band shifts deeper while the conduction band upper. (b) The band gap $E_G$ expands with pressure being dominated by the polarization of the lone pairs and the entrapment of the bonding pairs upon compression.

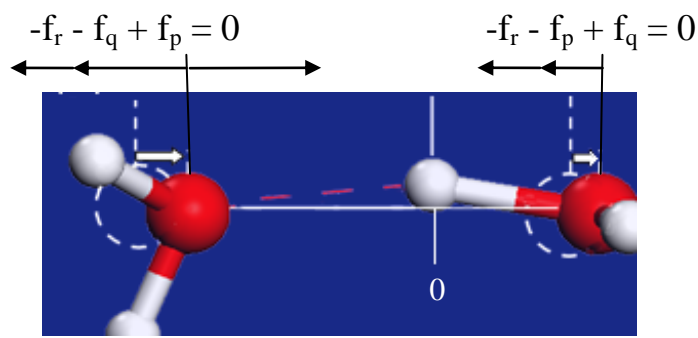

Figure 1

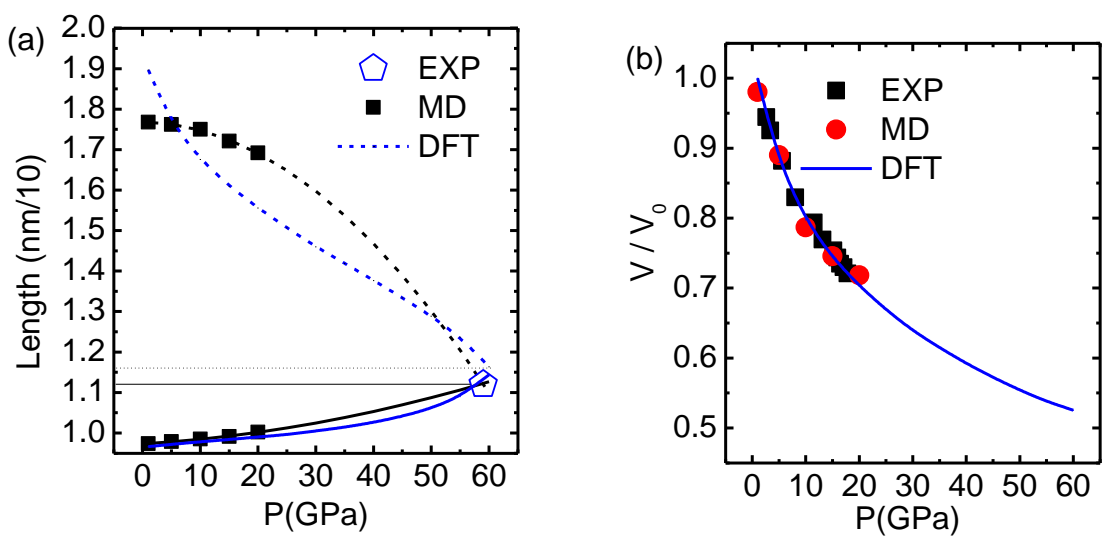

Figure 2

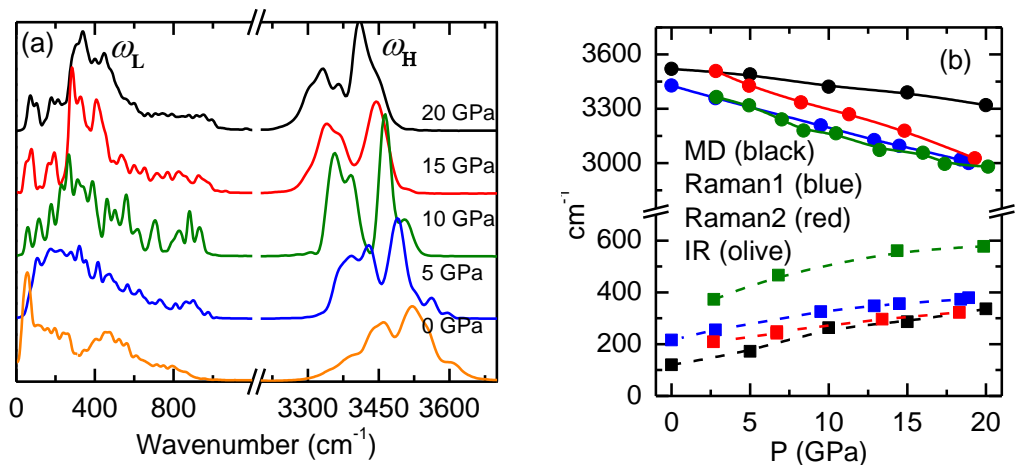

Figure 3

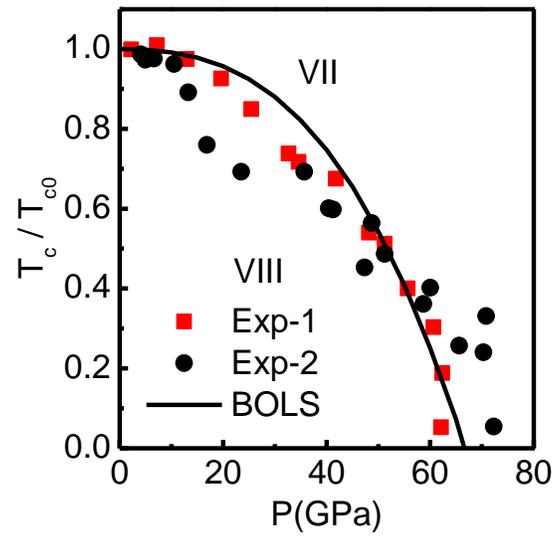

Figure 4

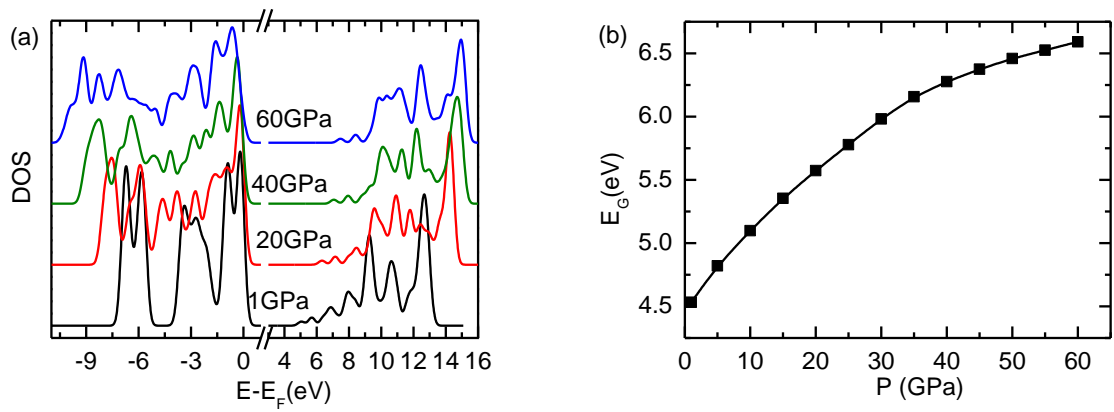

Figure 5